\documentclass[aps,prb,onecolumn,superscriptaddress]{revtex4-2}
\usepackage{epsfig,amssymb,amsfonts}
\usepackage{amsmath,amsfonts,amssymb}
\usepackage{graphicx}
\usepackage{caption}
\usepackage{xcolor}
\input{epsf}
\usepackage{epstopdf}
\usepackage{natbib}
\usepackage{braket}

\usepackage[labelformat=simple]{subcaption}
\makeatletter
\renewcommand\p@subfigure{\thefigure}
\makeatother

\begin{document}

\title{Efficient population transfer in a quantum dot exciton under phonon-induced decoherence via shortcuts to adiabaticity}

\author{Spyridon G. Kosionis}
\affiliation{Materials Science Department, School of Natural Sciences, University of Patras, Patras 26504, Greece}

\author{Sutirtha Biswas}
\affiliation{Department of Electrical and Computer Engineering, School of Engineering, Democritus University of Thrace, Xanthi 67100, Greece}

\author{Christina Fouseki}
\affiliation{Physics Department, School of Natural Sciences, University of Patras, Patras 26504, Greece}

\author{Dionisis Stefanatos}
\email{dionisis@post.harvard.edu}
\affiliation{Materials Science Department, School of Natural Sciences, University of Patras, Patras 26504, Greece}

\author{Emmanuel Paspalakis}
\affiliation{Materials Science Department, School of Natural Sciences, University of Patras, Patras 26504, Greece}
\email{paspalak@upatras.gr}

\begin{abstract}
In the present study, we apply shortcut to adiabaticity pulses (time-dependent Rabi frequency and detuning) for the efficient population transfer from the ground to the exciton state in a GaAs/InGaAs quantum dot with phonon-induced dephasing.
We use the time-evolving matrix product operator (TEMPO) method to propagate system in time and find that, for temperatures below $ 20 \ \text{K} $ and pulse duration up to $ 10 \ \text{ps} $, a very good transfer efficiency is obtained in general. We explain these results using a Bloch-like equation derived from a generalized Lindblad equation, which adequately describes system dynamics at lower temperatures. For higher temperatures, the transfer efficiency is significantly reduced except for subpicosecond pulses, where the shortcut Rabi frequency reduces to a delta pulse attaining a fast population inversion. The present work is expected to find application in quantum technologies which exploit quantum dots for single-photon generation on demand.
\end{abstract}

\keywords{acoustic phonons, Bloch equations, density matrix, excitons, Lindblad master equation, markovian dynamics, shortcuts to adiabaticity}

\maketitle

\section{INTRODUCTION}
\label{sec:intro}

Semiconductor quantum dots provide one of the most promising platforms for single-photon generation on demand, which is indispensable in quantum technologies \cite{Michler17,Senellart17}. Their discrete energy levels enable the emission of single photons with high purity and tunability. Also, quantum dots can be integrated into solid-state devices and photonic circuits, making them attractive for scalable quantum communication, quantum computing, and quantum cryptography. In addition, their compatibility with semiconductor fabrication techniques opens the door to large-scale integration and hybrid systems.

A stepping stone to exploit the full potential provided by these solid state systems for quantum information processing is the efficient controlled population transfer from the ground to the exciton state, despite the presence of decoherence and dissipation. This ensures that each excitation pulse results in the emission of exactly one photon, thus maximizing brightness and enabling deterministic operation. Achieving this requires precise quantum control strategies, which minimize unwanted multi-photon generation and timing jitter. Efficient exciton preparation is also crucial for generating photons with high indistinguishability, which is also an essential property in many quantum protocols.

In modeling the effect of the environment on the ground to exciton population transfer, two approaches are usually followed. According to the first approach, found mainly in earlier works, decay rates are inserted phenomenologically in the density matrix equations of the quantum dot \cite{Zrenner02,Stievater01,Kamada01,Htoon02,Wang05,Stufler05,Yu11,Schmigdall10,Simon11,Mukherjee20}, corresponding to a Markovian dynamics. In the more sophisticated approach, theoretical analysis \cite{Forstner03,Machnikowski04,Krugel05,Vagov07,Nazir08,Glassl11,Cutcheon10,Cutcheon11,Debnath12,Eastham13,Nazir16,Luker19} supported by experimental confirmation \cite{Ramsay10,Ramsay10b,Ramsey11} suggests that the major dephasing mechanism in the quantum dot system is the coupling to acoustic phonons \cite{Luker12,Reiter12,Wei14a,Mathew14a,Kaldewey17b,Ramachandran21,Kappe24a}, leading to a non-Markovian evolution. Despite the approach used to model decoherence, coherent control methods have been exploited to successfully achieve the desired population inversion. These include the application of resonant pulses \cite{Zrenner02,Stievater01,Kamada01,Htoon02,Wang05,Stufler05,Reiter12}, which excite Rabi oscillations in the quantum dot qubit driving it quickly to the target state but it is sensitive to experimental imperfections, as well as rapid adiabatic passage \cite{Debnath12,Eastham13,Schmigdall10,Yu11,Simon11,Luker12,Reiter12,Wei14a,Mathew14a,Kaldewey17b,Mukherjee20,Ramachandran21}, which robustly drives the system from the ground to the exciton state along a slower adiabatic path.

To accelerate the slow adiabatic dynamics while maintaining the robustness feature to some extent, a variety of methods called shortcuts to adiabaticity (STA) \cite{Odelin19} were developed, providing faster paths to the same target states. A basic problem explored using these methods is the development of efficient population transfer in two-level quantum systems \cite{Demirplak03,Berry09,Chen10a,Chen11,Ibanez12,Bason12a,Malossi13a,Ruschhaupt12a,Daems13a,Kiely14a,Martinez15a,Huang17a,Stefanatos19,Stefanatos19a}. Following the lines of the research discussed in the previous paragraph, in the present work we utilize the transitionless quantum driving shortcut method \cite{Demirplak03,Berry09}, developed for suppressing non-adiabatic transitions in fast quantum dynamics, to obtain pulses which are then used for the efficient ground to exciton population transfer in a GaAs/InGaAs quantum dot, under the presence of phonon-induced decoherence.  By applying the shortcut pulses to the quantum dot two-level system, as a time-dependent Rabi frequency and detuning,
we find using the time-evolving matrix product operator (TEMPO) method \cite{TEMPO-paper,PTTEMPO} that, for temperatures below $ 20 \ \text{K} $ and pulse duration up to $ 10 \ \text{ps} $, a very good transfer efficiency is obtained in general. For the detailed explanation of the results we employ a set of Bloch-like equations obtained from a generalized Lindblad equation \cite{Eastham13}, which can sufficiently describe the quantum dot dynamics at these lower temperatures. For higher temperatures, the transfer performance is considerably degraded except for pulses of subpicosecond duration, where the shortcut Rabi frequency essentially degenerates to a delta-like pulse achieving a fast population inversion.

The remainder of this paper is organized as follows. In Sec. \ref{sec:qdot}, we describe the system composed of a quantum dot interacting with a phonon bath and provide
the Bloch-like equations essentially capturing the quantum dot reduced dynamics for lower temperatures.
In Sec. \ref{sec:shortcut} we derive using
transitionless quantum driving the time-dependent Rabi frequency and detuning which perfectly prepare the exciton state in the absence of phonon-induced decoherence, while
in Sec. \ref{sec:results} we numerically study the effect of phonons on the population transfer efficiency,
for various values of pulse duration and temperature. Finally, Sec. \ref{sec:conclusion} summarizes our findings.

\section{QUANTUM DOT EXCITONS IN THE PRESENCE OF PHONON-INDUCED DECOHERENCE}
\label{sec:qdot}

Here, we consider a quantum dot driven at one of its discrete transition frequencies, specifically to its first excited state, by a circularly polarized laser pulse with time-varying amplitude and frequency
\begin{equation}
\mathcal{E}(t) = \mathcal{E}_0(t) \exp \left [ -i\omega_0t-i\phi(t)\right].
\end{equation}
Thus, we can model the quantum dot as a two-level system, with ground state $ |0\rangle $ and excited state the exciton state $ |X\rangle $. In the rotating wave approximation (RWA), within a rotating frame at the instantaneous driving frequency $ \omega_d(t) $, the system Hamiltonian can be expressed in terms of pseudospin operators $ \hat{s}_\alpha = \hat{\sigma}_\alpha/2, \ \alpha = x, y, z $, as
\begin{equation}
\label{Hdot}
\hat{H}_{\text{dot}} = \hbar \Delta(t) \hat{s}_z - \hbar \Omega(t) \hat{s}_x,
\end{equation}
where $ \Delta(t) = \dot{\phi}(t)$ is the field detuning from the exciton transition frequency $ \omega_0 $ and $ \Omega(t) = \mathcal{E}_0(t)\mu/\hbar$ is the Rabi frequency, with $\mathcal{E}_0(t)$ expressing the envelope of the applied field and $ \mu $ the dipole matrix element of the transition.
We should clarify that the RWA holds accurately when the decay and dephasing rates are negligible compared to the energy splitting. We also note that the population of the exciton state is expressed with respect to $ s_z = \langle \hat{s}_z \rangle $ as $ 1/2 + s_z $.

Based on the findings from experimental studies \cite{Ramsay10,Ramsay10b,Ramsey11}, it is reasonable to assume that the primary dephasing mechanism in quantum dots is the coupling to acoustic phonons, which is described via the interaction Hamiltonian
\begin{equation}
\label{coupling}
\hat{H}_C = \hat{s}_z \sum_q (g_q \hat{b}_q + g_q^* \hat{b}_q^\dagger),
\end{equation}
where $ \hat{b}_q $ and $ \hat{b}_q^\dagger $ represent the annihilation and creation operators for the acoustic phonon mode with wavevector $ q $ and $ g_q $ denotes the exciton-phonon coupling matrix element, that depends on the wavevector $q$ and is typically proportional to the deformation potential and the phonon energy. The phonon bath is described by the spectral density function $ J(\omega) = \sum_q |g_q|^2 \delta(\omega - \omega_q) $, which represents the energy distribution of phonon modes interacting with the system, in a form suitable for GaAs/InGaAs quantum dots \cite{Nazir16,Luker19,Ramsay10b}:
\begin{equation}
\label{sdensity}
J(\omega) = \frac{\hbar A}{\pi k_B} \omega^3 e^{-\omega^2 / \omega_c^2},
\end{equation}
with $ A = 11.2 \ \text{fs/K} $ and $ \hbar \omega_c = 2 \ \text{meV} $. This spectral density reflects the $ \omega^3 $ dependence at low frequencies, which is due to the coupling to acoustic phonons and their density of states, while the exponential cutoff emerges due to the finite size of the quantum dot. For a detailed derivation of the above spectral density, see Refs.\ \cite{Nazir16,Luker19}.
Note that, typically in the literature the quantum dot is coupled to acoustic phonons through the single exciton state $\ket{X}$ and the corresponding operator $\ket{X}\bra{X}$ appears in the coupling Hamiltonian, while here we use $\hat{s}_z=\hat{\sigma}_z/2=(\ket{X}\bra{X}-\ket{0}\bra{0})/2$, which also involves the ground state $\ket{0}$. From the completeness relation for the identity operator $I=\ket{X}\bra{X}+\ket{0}\bra{0}$ we find that the two operators are related by $\ket{X}\bra{X}=\hat{s}_z+I/2$. Since the identity operator does not couple the quantum dot to the phonons, the two approaches are equivalent. Also note that in Ref. \cite{Nazir16} the quantum dot - acoustic phonon coupling initially involves a linear combination of operators $\ket{X}\bra{X}$ and $\ket{0}\bra{0}$, but subsequently the latter is eliminated by subtracting a quantity proportional to the identity operator, leaving only the $\ket{X}\bra{X}$ term.

In Sec. \ref{sec:shortcut}, we present controls $\Omega(t)$ and $\Delta(t)$, obtained with the method of STA, which achieve efficient population transfer from the ground to the exciton state of the quantum dot described by Hamiltonian (\ref{Hdot}), even for arbitrarily small pulse durations. In the subsequent Sec. \ref{sec:results}, we numerically test the performance of these shortcut controls in the presence of the acoustic phonon dephasing mechanism, using the time-evolving matrix product operator (TEMPO) method \cite{TEMPO-paper}, for various values of the bath temperature and pulse duration. Introduced by Strathearn et al, TEMPO \cite{TEMPO-paper} is a numerically exact, yet efficient, method for modeling non-Markovian quantum dynamics mainly for treating cases where the coupling to the bath is diagonal in the system operators and the bath is homogeneous (i.e. the coupling of the quantum particle to the bath does not depend on the position of that particle), thus mostly for Caldeira–Leggett type of systems \cite{Caldeira83,Leggett87}, as the one studied here. The TEMPO method is based on a decomposition into a matrix product, which relies on the fact that two-time correlations in system parameters die off to become negligible at long times. This approach allows one to skip all analytical calculations related to the system-bath interactions and requires only the description of system Hamiltonian and the initial bath parameters.
To compute the quantum dot dynamics using the TEMPO algorithm, we have used the open source library OQuPy \cite{OQuPy-paper, OQuPy-code-v0.5.0} that exploits the process tensor formalism at its backend, which captures the complete influence of the environment on the system.

To gain some understanding regarding these numerical results, we use a set of equations
describing the time evolution of the Bloch vector components for the two-level system
(\ref{Hdot}) in the presence of the acoustic phonon dephasing mechanism, derived by Eastham
et al. \cite{Eastham13}. Recall that the density matrix of a quantum system is positive semi-
definite, as its eigenvalues are the occupation probabilities of the corresponding eigenstates. For
open quantum systems, the state of the basic quantum system is characterized by the reduced
density matrix, which is produced by tracing out the degrees of freedom of the environment
\cite{Breuer02,Rivas11}. Under the Born-Markov approximation, the time evolution of the
reduced density matrix is given by a Markovian master equation and the most common
Markovian master equation is the Lindblad master equation, which preserves positivity
\cite{Lindblad76}. Eastham et al. implemented the approach presented in Refs.
\cite{Ramsay10b,Ramsey11} to derive a Lindblad master equation with time-dependent
damping and decoherence that preserves positivity and governs the dynamics of the reduced
density matrix of the two-level quantum dot system interacting with a phonon environment.

This equation is derived in the limit where $ \Delta(t) $ and $ \Omega(t) $ change slowly with
time, thus they can be treated effectively as constants, in which case the effect of acoustic
phonons is obtained by transforming the composite system to the interaction picture and then
applying the Born-Markov approximation. This procedure additionally demands that the system
state remains approximately constant over the phonon bath correlation time ($ \sim \omega_c^{-
1} $), which is equivalent to the condition that the bath spectral density (\ref{sdensity}) does not
change appreciably over the system effective linewidth. This implies that both the decay rates
and the rates of change of $ \Delta(t) $ and $ \Omega(t) $, contributing to the linewidth, should
be kept small. Under the above assumptions, the phonon degrees of freedom can be traced out
following the standard procedure, leaving an equation for the reduced density matrix of the
quantum dot. Eastham et al. \cite{Eastham13} proceed further to the secularization of the
derived equation (averaging over a time short compared to decay rates but long relative to the
time scales of the system Hamiltonian \cite{Dumcke79}), and this procedure yields the
generalized Lindblad master equation.

Using this equation, the time evolution of the Bloch vector components of the quantum dot two-level system, $ \mathbf{s} = \begin{pmatrix} s_x & s_y & s_z \end{pmatrix}^T $ with $ s_\alpha = \langle \hat{s}_\alpha \rangle $ for $ \alpha = x, y, z $, is found to be described by the following set of equations \cite{Eastham13}:
\begin{subequations}
\label{bloch}
\begin{eqnarray}
\dot{s}_x &=& -\frac{\Omega}{2\Lambda}(\gamma_a-\gamma_e)-\frac{\Delta^2+2\Omega^2}{2\Lambda^2}(\gamma_a+\gamma_e)s_x \nonumber \\
          & & - \Delta s_y + \frac{\Delta\Omega}{2\Lambda^2}(\gamma_a+\gamma_e)s_z, \label{sx} \\
\dot{s}_y &=& \Delta s_x -\frac{\gamma_a+\gamma_e}{2}s_y + \Omega s_z, \label{sy} \\
\dot{s}_z &=&  \frac{\Delta}{2\Lambda}(\gamma_a-\gamma_e) + \frac{\Delta\Omega}{2\Lambda^2}(\gamma_a+\gamma_e)s_x - \Omega s_y\nonumber \\
          & &  -\frac{2\Delta^2+\Omega^2}{2\Lambda^2}(\gamma_a+\gamma_e)s_z, \label{sz}
\end{eqnarray}
\end{subequations}
with
\begin{equation}
\label{Lambda}
\Lambda = \sqrt{\Omega^2 + \Delta^2}
\end{equation}
and the phonon absorption and emission rates are given by the expressions
\begin{subequations}
\label{decay_rates}
\begin{eqnarray}
\gamma_e &=& 2\left(\frac{\Omega}{2\Lambda}\right)^2\pi J(\Lambda) [1 + n_b(\Lambda)], \label{emission} \\
\gamma_a &=& 2\left(\frac{\Omega}{2\Lambda}\right)^2\pi J(\Lambda) n_b(\Lambda), \label{absorption}
\end{eqnarray}
\end{subequations}
where
\begin{equation}
\label{n_phonons}
n_b(\omega) = \frac{1}{e^{\frac{\hbar\omega}{k_B\Theta}}-1}
\end{equation}
denotes the phonon occupation number at frequency $ \omega $ and temperature $\Theta$. Observe that these rates depend on the applied fields and thus are time-dependent. As discussed in Ref. \cite{Eastham13}, for temperatures lying below 20 K, the peak damping rates and corresponding linewidths are not substantial compared to the width of the spectral function and thus the assumption that the value of this function is constant over that spectral range remains practically valid. For these temperatures, we will use the above equations to gain qualitative insight regarding the numerical results obtained with TEMPO.


\section{EFFICIENT EXCITON STATE PREPARATION VIA SHORTCUTS TO ADIABATICITY}
\label{sec:shortcut}

According to the STA methodology, see for example Ref. \cite{Stefanatos19}, to achieve highly efficient population transfer to the exciton state in the QD, even for short durations for which the adiabaticity condition is not fulfilled, a pulsed electromagnetic field with appropriate time-dependent Rabi frequency and detuning should be applied to the corresponding two-level system. The exact expressions for these control functions are derived by applying the transitionless quantum driving shortcut protocol. The basic idea behind this protocol is the elimination of the non-diagonal diabatic terms which appear in the Hamiltonian of the two-level system when expressed in the adiabatic basis. To address this, an anti-diabatic term is added to the Hamiltonian, which counteracts the diabatic transitions and ensures that the system evolves following the adiabatic eigenstates of the original Hamiltonian, even for very short pulses.


For example, consider that the following fields are applied to the original Hamiltonian (\ref{Hdot}):
\begin{subequations}
\label{nfields}
\begin{eqnarray}
\Delta_n(t) &=& E_0(t) \cos\theta(t), \\
\Omega_n(t) &=& E_0(t) \sin\theta(t),
\end{eqnarray}
\end{subequations}
parametrized by the time-dependent amplitude $E_0(t)$ and mixing angle $\theta(t)$, where the subscript refers to “nominal”. If the counterdiabatic term $\hat{H}_{\text{cd}} = -\dot{\theta} \hat{s}_y$ is added to the original Hamiltonian $\hat{H}_{\text{dot}}$, then, under the total Hamiltonian $\hat{H}_{\text{tot}} = \hat{H}_{\text{dot}} + \hat{H}_{\text{cd}}$, the system state can track the eigenstates of $\hat{H}_{\text{dot}}$:
\begin{subequations}
\label{eigenstates}
\begin{eqnarray}
|\psi_+\rangle &=& \begin{pmatrix} \cos\frac{\theta}{2} \\ -\sin\frac{\theta}{2} \end{pmatrix}, \\
|\psi_-\rangle &=& \begin{pmatrix} \sin\frac{\theta}{2} \\ \cos\frac{\theta}{2} \end{pmatrix},
\end{eqnarray}
\end{subequations}
even for very short pulses. The desired population inversion can be implemented along any of the eigenstates by choosing $\theta(t)$ to satisfy the appropriate boundary conditions. The problem with the above approach is that it needs the extra counter-diabatic term to work, proportional to $\hat{s}_y$. However, there is a way to achieve the desired population transfer with a Hamiltonian which has the form of $\hat{H}_{\text{dot}}$, without such a term.
As discussed in Ref. \cite{Stefanatos19}, if we make the following unitary transformation
\begin{equation}
\hat{U}(t) = e^{-i\beta(t) \hat{s}_z},
\end{equation}
corresponding to a rotation around $z$-axis, the transformed total Hamiltonian is
\begin{equation}
\label{Ht}
\hat{U}^\dagger \hat{H}_{\text{tot}} \hat{U} - i\hbar\hat{U}^\dagger \dot{\hat{U}} = \hbar(E_0\cos\theta-\dot{\beta})\hat{s}_z+\hbar(E_0\sin\theta\cos \beta+\dot{\theta}\sin \beta)\hat{s}_x
           +\hbar(\dot{\theta}\cos \beta-E_0\sin\theta\sin \beta)\hat{s}_y.
\end{equation}
Furthermore, if we choose
\begin{equation}
\beta = \arctan\left(\frac{\dot{\theta}}{E_0 \sin\theta}\right)
\end{equation}
then the $\hat{s}_y$ term in Eq. (\ref{Ht}) is eliminated
and the transformed state $ \hat{U}^\dagger(t) |\psi(t)\rangle $ evolves under the transformed total Hamiltonian which has the original form of $ \hat{H}_{\text{dot}} $ but with Rabi frequency and detuning given in the following equations:
\begin{subequations}
\label{fields}
\begin{eqnarray}
\Omega(t) &=& E_0 \sin\theta \cos\beta + \dot{\theta} \sin\beta \label{omega} \\
          &=& \sqrt{E_0^2 \sin^2\theta + \dot{\theta}^2}, \nonumber \\
\Delta(t) &=& E_0 \cos\theta - \dot{\beta} \label{delta} \\
          &=& \frac{E_0^3 \sin^2\theta \cos\theta + \dot{E}_0 \dot\theta \sin\theta + E_0 (2\dot{\theta}^2 \cos\theta - \ddot{\theta} \sin\theta)}{E_0^2 \sin^2\theta + \dot{\theta}^2}, \nonumber
\end{eqnarray}
\end{subequations}
The time-dependent functions $ E_0(t) $ and $ \theta(t) $, which essentially determine the applied Rabi frequency $ \Omega(t) $ and detuning $ \Delta(t) $, are derived by imposing the appropriate boundary conditions and then making a polynomial interpolation over the time interval where the pulses are applied, from 0 to the final time $ T $. The boundary conditions ensure that the population transfer in the transformed state takes place along one of the eigenstates ($ |\psi_-\rangle $), the additional counter-diabatic term in the Hamiltonian vanishes at the boundary times, and the control functions are smooth enough and vanish at the boundary points. The procedure is described in detail in Ref. \cite{Stefanatos19}, where the following expressions are derived for driving the system along eigenstate $ |\psi_-\rangle $:
\begin{equation}
E_0(\tau) = \Omega_0 \tau (1 - \tau), \label{E}
\end{equation}
and
\begin{equation}
\theta(\tau) = \pi \tau^2 (3 - 2\tau).  \label{theta}
\end{equation}

\begin{figure}[t]
 \centering
 \vspace{-2cm}
\begin{subfigure}[t!]{0.4\textwidth}
    \centering\vspace{0cm}\caption{}\includegraphics[width=\linewidth]{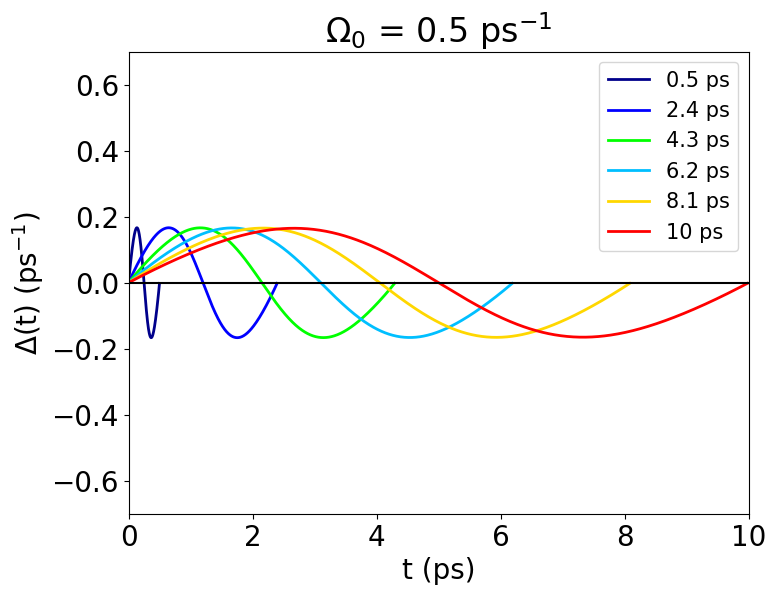}\label{fig:D05}
\end{subfigure}
\begin{subfigure}[c]{0.4\textwidth}
    \centering\vspace{0cm}\caption{}\includegraphics[width=\linewidth]{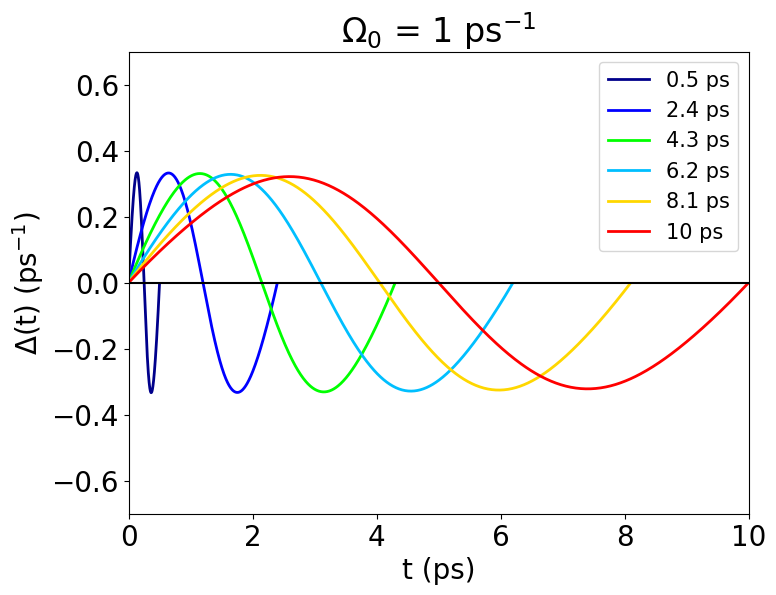}\label{fig:D1}
\end{subfigure} \\
\begin{subfigure}[t!]{0.4\textwidth}
    \centering\vspace{0cm}\caption{}\includegraphics[width=\linewidth]{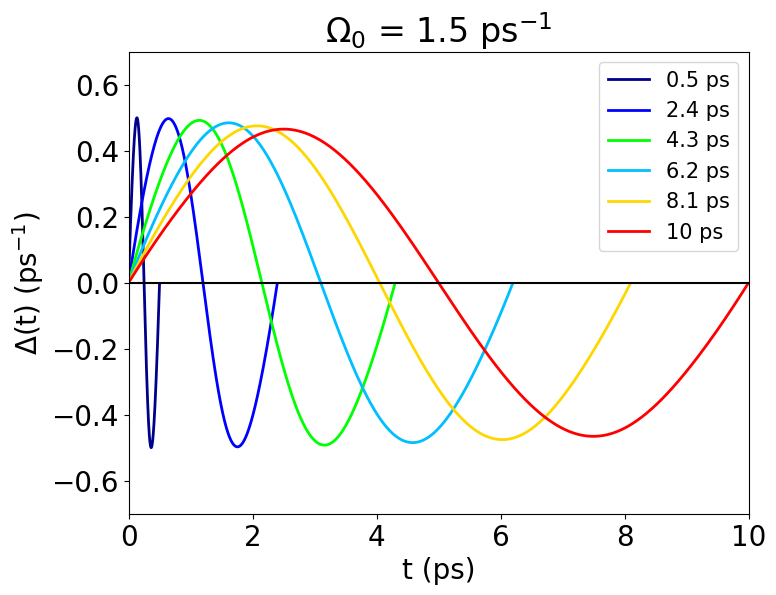}\label{fig:D15}
\end{subfigure}
\begin{subfigure}[c]{0.4\textwidth}
    \centering\vspace{0cm}\caption{}\includegraphics[width=\linewidth]{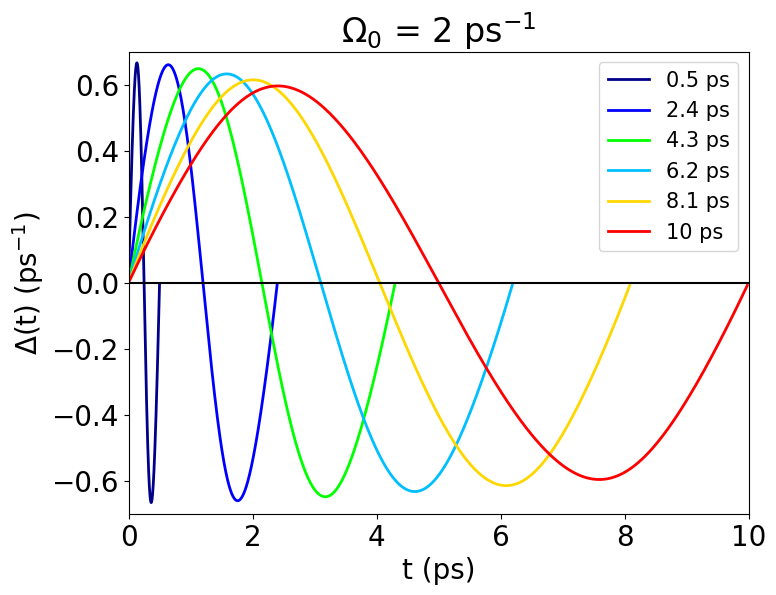}\label{fig:D2}
\end{subfigure}
\caption{Time-dependent detuning $ \Delta(t) $ of the applied field obtained from Eqs. (\ref{fields}) using Eqs. (\ref{E}) and (\ref{theta}), for various values of the pulse duration $T = 0.5, 2.4, 4.3, 6.2, 8.1, 10 \ \text{ps}$ and parameter $ \Omega_0 = 0.5 \ \text{ps}^{-1} $ (a), $ 1 \ \text{ps}^{-1} $ (b), $ 1.5 \ \text{ps}^{-1} $ (c), and $ 2 \ \text{ps}^{-1} $ (d).}
\label{fig:controls}
\end{figure}
\begin{figure}[h]
    \centering
   \includegraphics[width=0.5\textwidth]{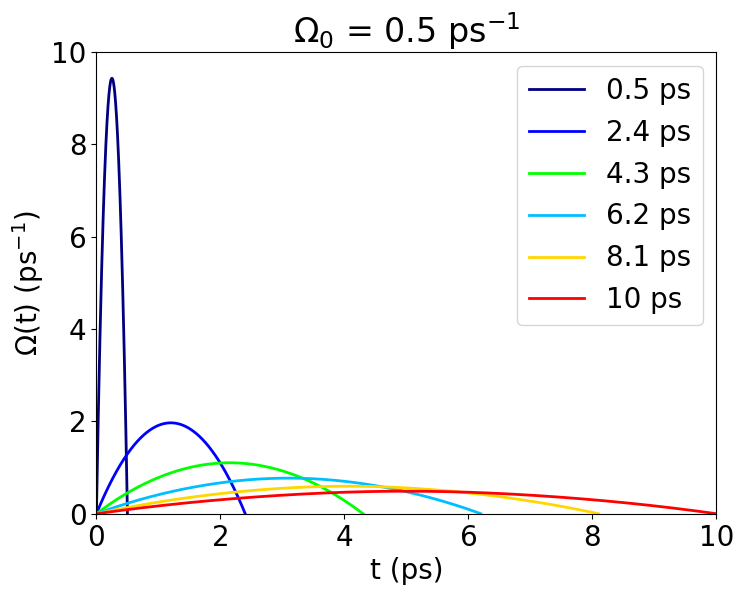} 
   \vspace{-0cm}
    \caption{Time-dependent Rabi frequency $ \Omega(t) $ obtained as in Fig. 1, where note that, since there is no visible difference for different values $ \Omega_0 = 0.5, 1, 1.5, 2 \ \text{ps}^{-1} $,
    only the case with $ \Omega_0 = 0.5 \ \text{ps}^{-1} $ is shown.}
    \label{fig:W}
\end{figure}

Here, we have introduced for simplicity the normalized time $ \tau = t/T $, with $ T $ the pulse duration, but emphasize that the derivatives in Eqs. (\ref{fields}) are calculated with respect to the actual time $ t $, not the normalized time $ \tau $. $ \Omega_0 $ is a constant amplitude parameter. 
As explained in Ref. \cite{Stefanatos19}, the controls $ \Omega(t) $ and $ \Delta(t) $ given in Eqs. (\ref{fields}), calculated with functions $ E_0(t) $ and $ \theta(t) $ given in Eqs. (\ref{E}) and (\ref{theta}), achieve perfect population inversion also for the original state (not only the transformed one), from the ground to the exciton state. In Fig. \ref{fig:controls}, we depict the detuning $ \Delta(t) $ for various values of the pulse duration $T = 0.5, 2.4, 4.3, 6.2, 8.1, 10 \ \text{ps}$ and parameter $ \Omega_0 = 0.5 \ \text{ps}^{-1} $ (a), $ 1 \ \text{ps}^{-1} $ (b), $ 1.5 \ \text{ps}^{-1} $ (c), and $ 2 \ \text{ps}^{-1} $ (d). The corresponding Rabi amplitude $ \Omega(t) $ does not present visible differences for the values of $\Omega_0$ used, thus only the case with $ \Omega_0 = 0.5 \ \text{ps}^{-1} $ is shown in Fig. \ref{fig:W}. Note that shortcut pulses have been implemented in experiments in various platforms, see Ref. \cite{Odelin19} for a list up to 2019, while many more have followed.


In the next section, we investigate numerically the effect of acoustic phonons on the population transfer to the exciton state when the shortcut controls are applied to the system, over a range of temperatures and pulse durations.

\section{EFFECT OF ACOUSTIC PHONONS ON THE GROUND-TO-EXCITON POPULATION TRANSFER}
\label{sec:results}

\begin{figure*}[t]
 \centering
 \vspace{-1.5cm}
 \begin{subfigure}[t!]{0.49\textwidth}
    \centering\caption{}\includegraphics[width=\linewidth]{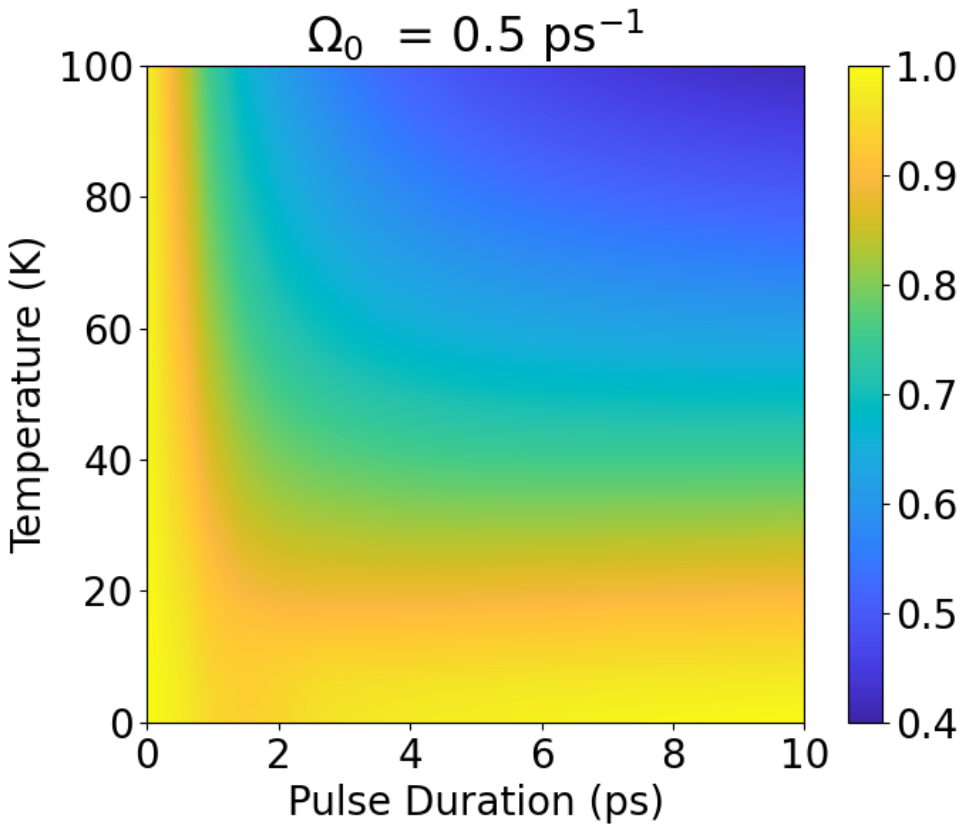}\label{fig:L05}
\end{subfigure}
\begin{subfigure}[t!]{0.49\textwidth}
    \centering\caption{}\includegraphics[width=\linewidth]{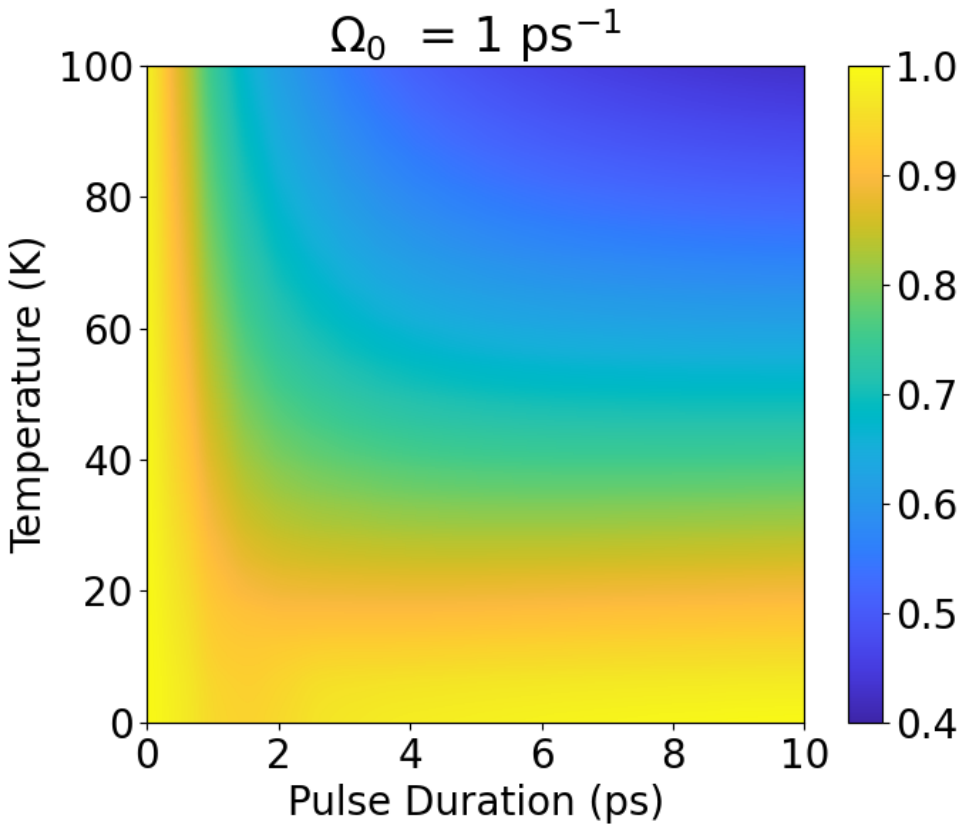}\label{fig:L1}
\end{subfigure} \\
\vspace{-4cm}
\begin{subfigure}[c]{0.49\textwidth}
    \centering\caption{}\includegraphics[width=\linewidth]{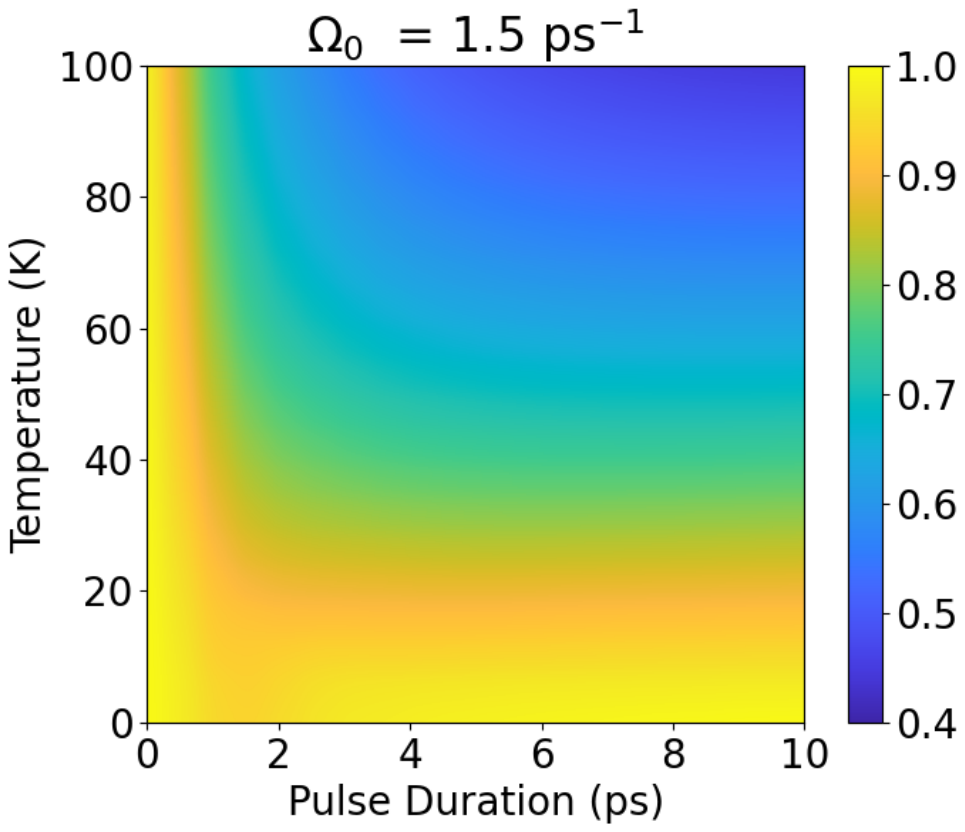}\label{fig:L15}
\end{subfigure}
\begin{subfigure}[c]{0.49\textwidth}
    \centering\caption{}\includegraphics[width=\linewidth]{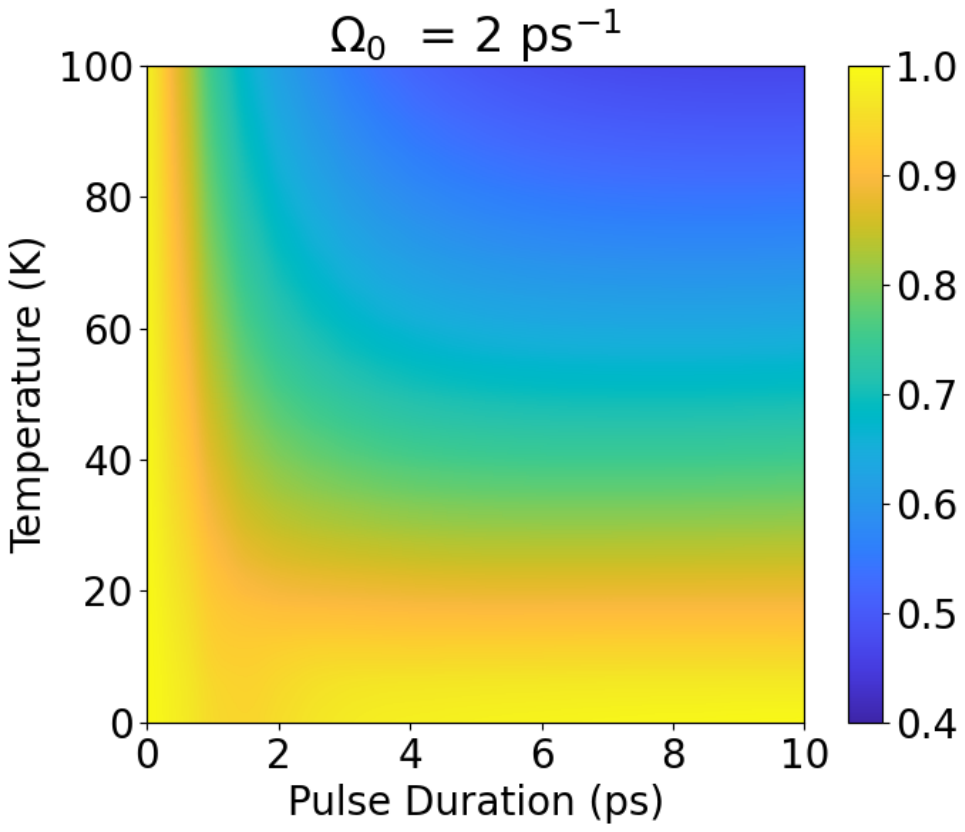}\label{fig:L2}
\end{subfigure}
\vspace{-3.5cm}
\caption{Population of exciton state after the application of shortcut pulses, as a function of pulse duration up to $ 10 \ \text{ps} $ and temperature in the range $ 0-100 \ \text{K} $, for different values of parameter $ \Omega_0 = 0.5 \ \text{ps}^{-1} $ (a), $ 1 \ \text{ps}^{-1} $ (b), $ 1.5 \ \text{ps}^{-1} $ (c), and $ 2 \ \text{ps}^{-1} $ (d). The results are obtained using the TEMPO method.}
\label{fig:LHTD}
\end{figure*}

In Fig. \ref{fig:LHTD},  we display the population of the exciton state obtained with TEMPO after the application of the shortcut pulses, calculated from Eq. (\ref{fields}) with four different values of parameter $\Omega_0$ (0.5, 1, 1.5 and 2 $\text{ps}^{-1}$), as a function of pulse duration up to $ 10 \ \text{ps} $ and for temperatures in the range $0-100 \ \text{K}$. For a detailed discussion of the TEMPO convergence, see Appendix. We observe that for temperatures up to $ 20 \ \text{K} $ and shorter pulses, with duration below $ 1 \ \text{ps} $, a high transfer efficiency is achieved. The performance is degraded for pulses with duration in the range around  $ 1-2 \ \text{ps} $, while it is improved for longer durations, especially for temperatures below $ 10 \ \text{K} $. The good performance of subpicosecond pulses is preserved even for higher temperatures, although for shorter pulse duration with increasing temperature. We also observe that in general the transfer efficiency worsens with increasing temperature and duration, especially for temperatures higher than  $ 30 \ \text{K} $, while the contour plots acquire a characteristic hyperbolic shape. Finally, the performance is robust against parameter $\Omega_0$, for the values used.

\begin{figure*}[t]
 \centering
 \vspace{-1.5cm}
 \begin{subfigure}[t!]{0.49\textwidth}
    \centering\caption{}\includegraphics[width=\linewidth]{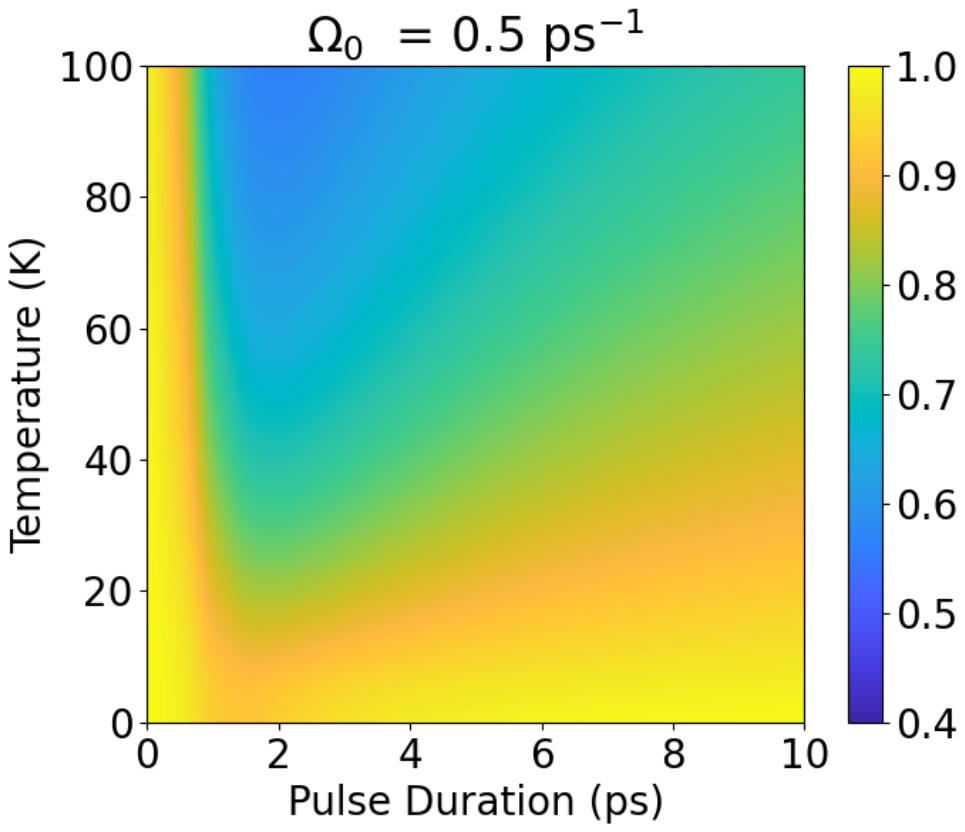}\label{fig:B05}
\end{subfigure}
\begin{subfigure}[t!]{0.49\textwidth}
    \centering\caption{}\includegraphics[width=\linewidth]{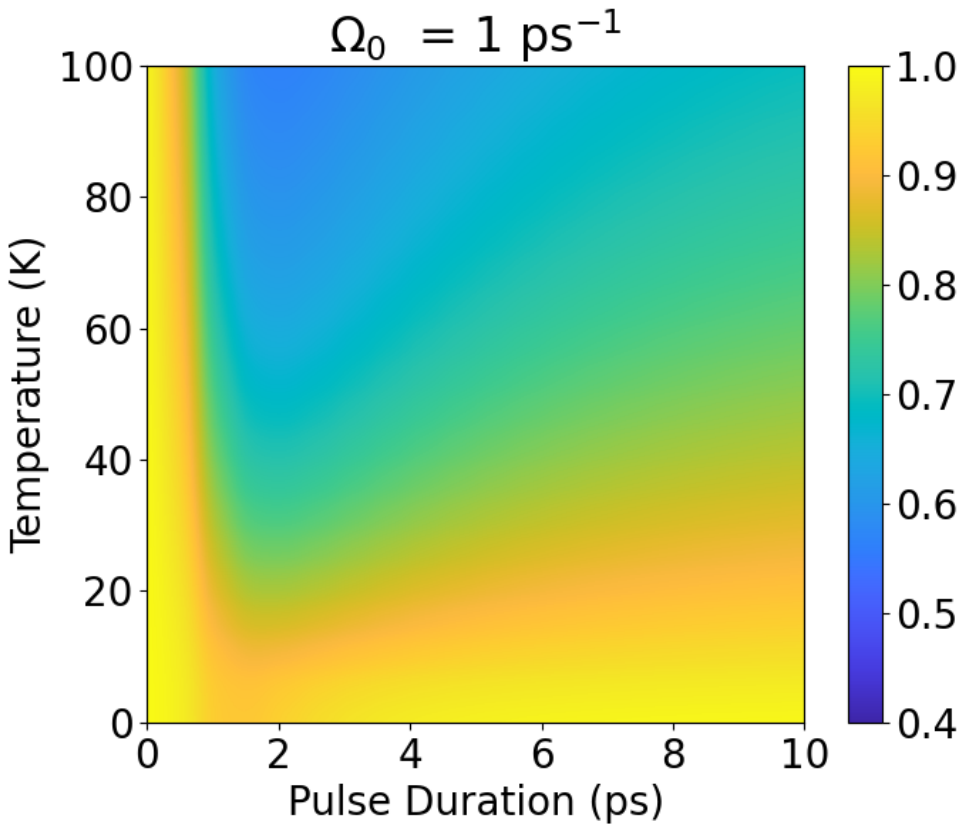}\label{fig:B1}
\end{subfigure} \\
\vspace{-4cm}
\begin{subfigure}[c]{0.49\textwidth}
    \centering\caption{}\includegraphics[width=\linewidth]{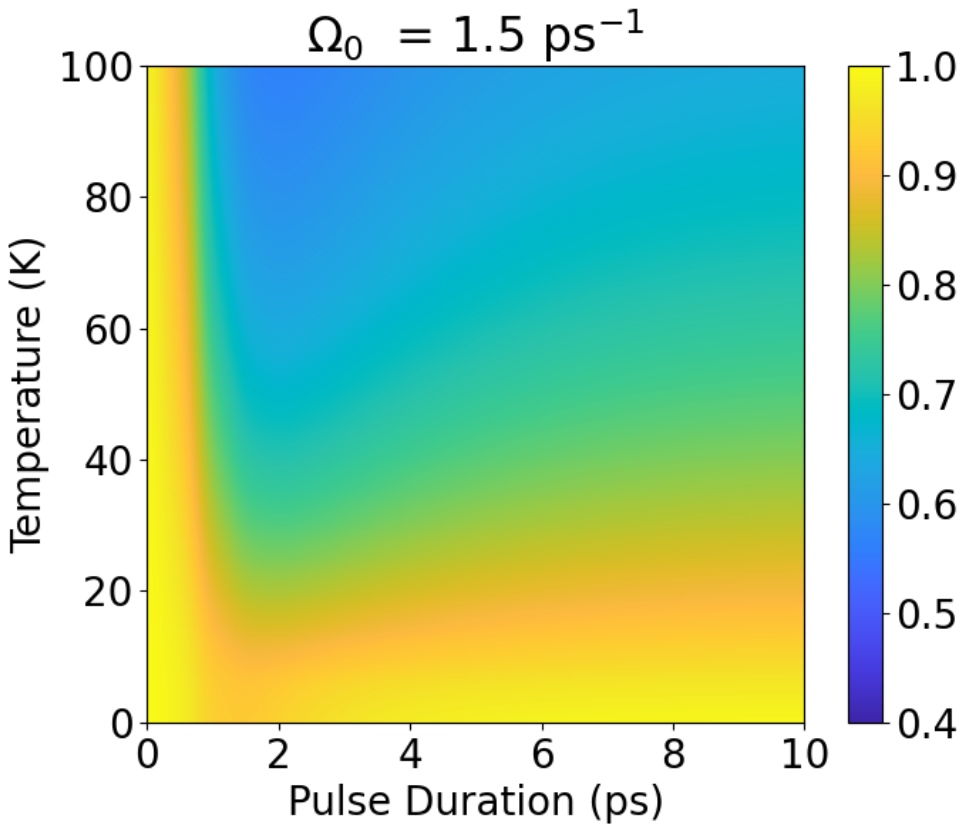}\label{fig:B15}
\end{subfigure}
\begin{subfigure}[c]{0.49\textwidth}
    \centering\caption{}\includegraphics[width=\linewidth]{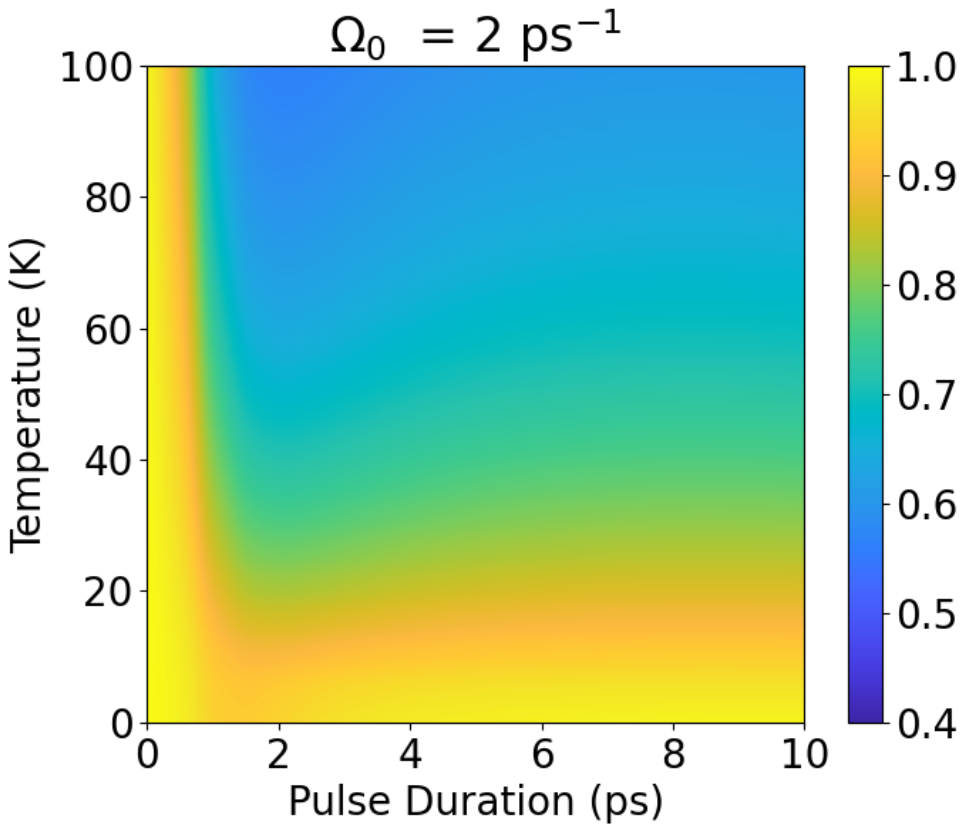}\label{fig:B2}
\end{subfigure}
\vspace{-3.5cm}
\caption{Same as in Fig. \ref{fig:LHTD} but the results are obtained using Eqs. (\ref{bloch}) instead of TEMPO.}
\label{fig:bloch}
\end{figure*}

\begin{figure}[t]
 \centering
\vspace{-2cm}
 \begin{subfigure}[t!]{0.6\textwidth}
    \centering\caption{}\includegraphics[width=\linewidth]{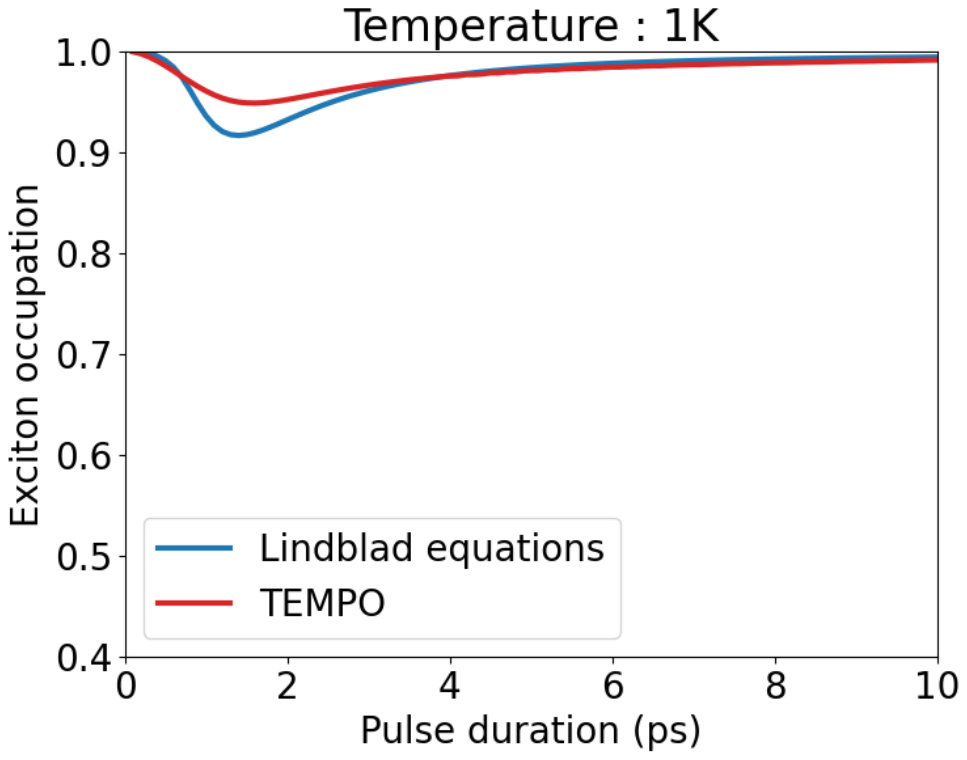}\label{fig:LT1}
\end{subfigure}  \\
\vspace{-6cm}
\begin{subfigure}[t!]{0.6\textwidth}
    \centering\caption{}\includegraphics[width=\linewidth]{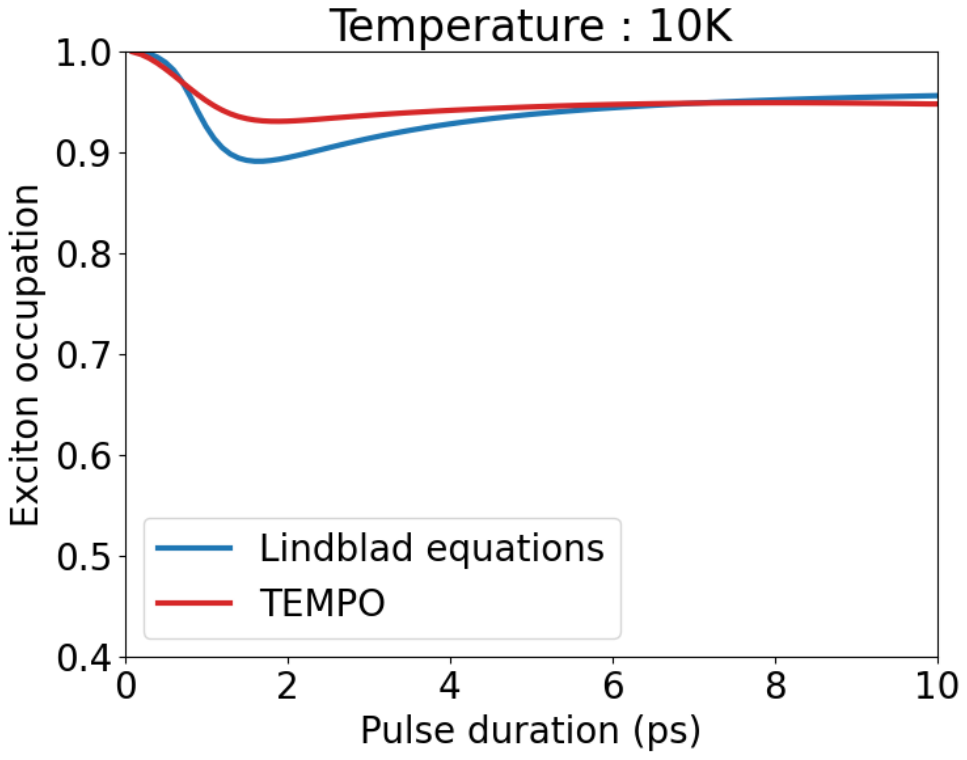}\label{fig:LT10}
\end{subfigure}  \\
\vspace{-6cm}
\begin{subfigure}[c]{0.6\textwidth}
    \centering\caption{}\includegraphics[width=\linewidth]{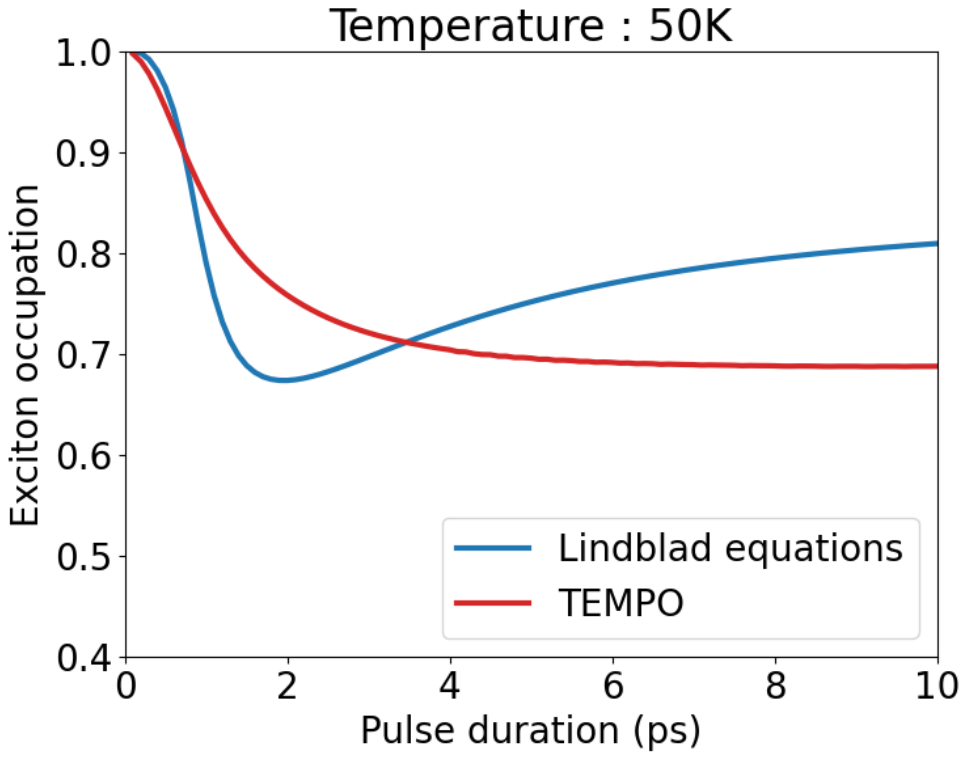}\label{fig:LT50}
\end{subfigure}
\vspace{-6cm}
\caption{Final exciton population achieved with the shortcut pulses versus pulse duration, calculated with TEMPO (red curve) and Eqs. (\ref{bloch}) (blue curve), for temperatures $ 1 \ \text{K} $ (a), $ 10 \ \text{K} $ (b), and $ 50 \ \text{K} $ (c).}
\label{fig:TL}
\end{figure}

To gain an understanding of the results obtained for lower temperatures, we will use Eqs. (\ref{bloch}). We first verify the validity of the equations in this range of temperatures. In
Fig. \ref{fig:bloch} we display the population of exciton state obtained after applying the shortcut control pulses in Eqs. (\ref{bloch}), as a function of pulse duration up to $ 10 \ \text{ps} $ and temperature in the range $ 0-100 \ \text{K} $, using the same values $\Omega_0$ for the controls as in Fig. \ref{fig:LHTD}. In comparison to the results of Fig. \ref{fig:LHTD}, we observe that model (\ref{bloch}) essentially captures the effect of phonons on the final exciton population for temperatures below $ 20 \ \text{K} $. This observation can be also verified in Figs. \ref{fig:LT1} and \ref{fig:LT10}, where we display the final exciton population obtained with TEMPO (red curve) and model (\ref{bloch}) (blue curve) versus pulse duration for the low temperatures $ 1 \ \text{K} $ and $ 10 \ \text{K} $, respectively.

\begin{figure*}[t]
 \centering
 \begin{subfigure}[t!]{0.49\textwidth}
    \centering\caption{}\includegraphics[width=\linewidth]{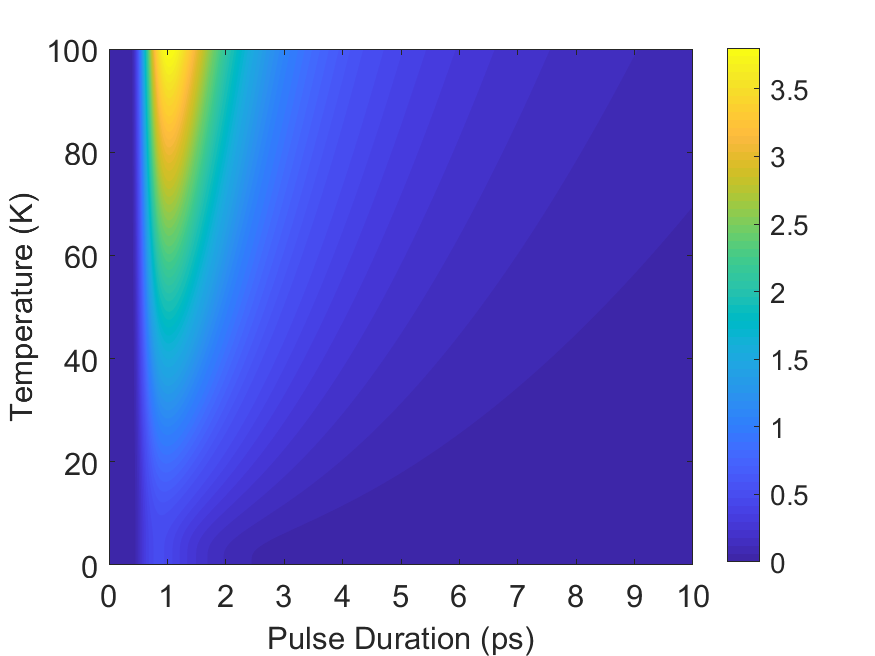}\label{fig:dec}
\end{subfigure}
\begin{subfigure}[t!]{0.49\textwidth}
    \centering\caption{}\includegraphics[width=\linewidth]{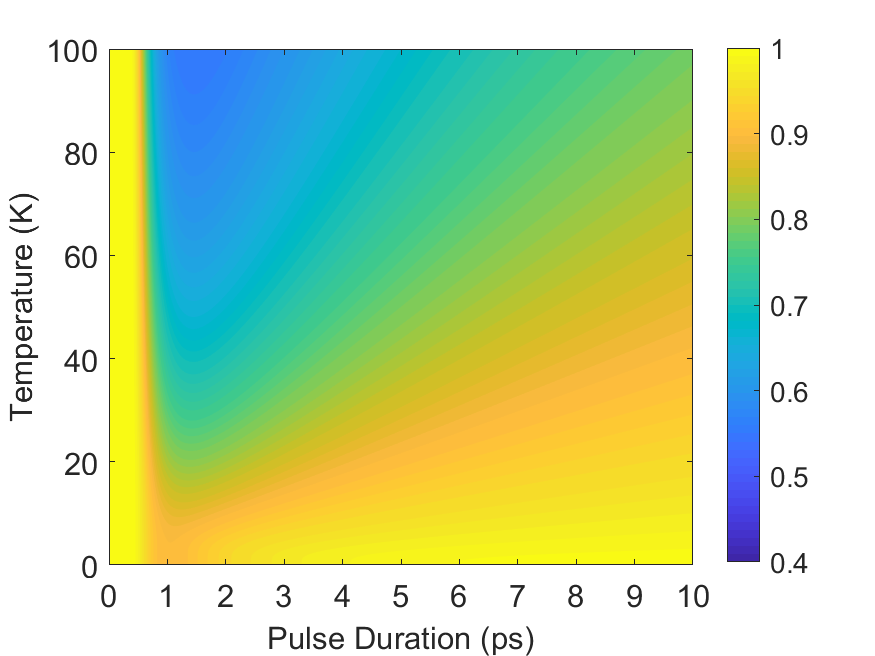}\label{fig:pipulse}
\end{subfigure}
\caption{(a) Normalized decay rate $(\Lambda/\Omega)^2(\gamma_a+\gamma_e)$ of model (\ref{bloch}) evaluated at $\Lambda=\pi/T$ and (b) Final exciton population obtained with model (\ref{bloch}) for a constant pulse $\Omega=\pi/T$, as a function of pulse duration $T$ up to $ 10 \ \text{ps} $ and temperature in the range $ 0-100 \ \text{K} $.}
\label{fig:decay}
\end{figure*}

Although from the above figures it appears that model (\ref{bloch}) rather overestimates the deteriorating effect of phonons for pulse durations around $ 1-2 \ \text{ps} $, it can still be exploited for drawing qualitative conclusions regarding the results obtained with TEMPO for lower temperatures. For example, the extremely good performance of the shortcut scheme for subpicosend pulses can be attributed to the fact that for such durations $\Omega(t)$ is essentially a delta pulse, as it is indicated by Fig. \ref{fig:W}, leading to a fast population inversion of $s_z$ mostly along $s_y$, without leaving much time for dissipative terms involving $\gamma_a+\gamma_e$ to act.
We next move to explain the degradation of performance for pulse duration around the range $ 1-2 \ \text{ps} $
and the subsequent improvement for longer duration, at least for temperatures below $ 10 \ \text{K} $, see for example Fig. \ref{fig:LT1}. For this reason, in Fig. \ref{fig:decay} we plot the normalized decay rate $(\Lambda/\Omega)^2(\gamma_a+\gamma_e)$ of model (\ref{bloch})
evaluated at $\Lambda=\pi/T$, as a function of duration $T$ and temperature. Observe that the decay obtains higher values in the aforementioned range around $ 1-2 \ \text{ps} $ while is reduced for longer pulse duration, capturing the behavior shown in Fig. \ref{fig:LHTD}. The degradation for higher temperatures is also present in the diagram.

For the case where $\Omega_0 = 0.5 \ \text{ps}^{-1}$, depicted in Fig. \ref{fig:L05}, we can actually move the analysis one step further. Observe from Figs. \ref{fig:D05} and \ref{fig:W} that in this case the shortcut detuning is rather small compared to the Rabi frequency. Then, we can simplify the analysis by setting $\Delta = 0$ in Eqs. (\ref{bloch}). Indeed, with this choice $s_x$ is decoupled from $s_y, s_z$, while the latter variables satisfy the simplified system
\begin{subequations}
\label{simplified}
\begin{eqnarray}
\dot{s}_y &=& -\frac{\gamma_a+\gamma_e}{2}s_y + \Omega s_z, \label{ssy} \\
\dot{s}_z &=&  - \Omega s_y  -\frac{\gamma_a+\gamma_e}{2}s_z, \label{ssz}
\end{eqnarray}
\end{subequations}
which can be integrated even for time-dependent Rabi frequency. Using the initial conditions $s_y(0)=0$ and $s_z(0)=-1/2$, expressing that the ground-state is initially fully populated, we find the following expression for the exciton population at the final time $t=T$
\begin{eqnarray}
\label{population}
P_e(T) &=& \frac{1}{2}+s_z(T) \nonumber \\
    &=& \frac{1}{2}\left[1-e^{-\frac{1}{2}\int_0^T(\gamma_a+\gamma_e)dt}\cos{\left(\int_0^T\Omega(t)dt\right)}\right],
\end{eqnarray}
where note that the decay rates $\gamma_a, \gamma_e$ are time-dependent through the Rabi frequency $\Omega(t)$. Now, instead of using the shortcut time-dependent $\Omega(t)$ in Eq. (\ref{population}), we take as before a constant $\pi$-pulse of duration $T$, with $\Omega=\pi/T$. This choice is justified at least for smaller $T$, since in this case the $\dot{\theta}$ term dominates Eq. (\ref{omega}), thus $\Omega(t)\approx\dot{\theta}(t)$ and $\int_0^T\Omega(t)dt\approx \theta(T) = \pi$. In Fig. \ref{fig:pipulse} we plot the corresponding exciton population from Eq. (\ref{population}) and observe that indeed captures the low temperature behavior observed in Figs. \ref{fig:L05} and \ref{fig:B05}, considering the rough approximations made.

For larger temperatures the model of Eq. (\ref{bloch}) essentially ceases to be valid, as also shown in Fig. \ref{fig:LT50}, with the exception of very short durations; however, we can still draw some interesting conclusions. First, in Fig. \ref{fig:LHTD} the very good performance for small duration persists for higher temperatures, indicating that the picture of a delta-like pulse quickly inverting the population is still relevant. Second, the hyperbolic-like contour plots observed in Fig. \ref{fig:LHTD} for temperatures larger than $ 30 \ \text{K} $ may be explained as follows. For higher temperatures, we expect that the losses in the system are mainly determined by the number of phonons, Eq. (\ref{n_phonons}). For $\hbar\Lambda/k_B\Theta\ll 1$ it is $n_b\approx k_B\Theta / \hbar\Lambda$. If we use as before $\Lambda=\pi/T$, where $T$ denotes the pulse duration, then we have $n_b\sim \Theta T$. The contours of constant phonon number, which dominates losses and thus determines the performance contours, are the hyperbolas $\Theta T=\text{constant}$ on the $T-\Theta$ space.

We close the analysis of the TEMPO results presented in Fig. \ref{fig:LHTD} by pointing out the robustness of the transfer with respect to parameter $\Omega_0$, which is used in the calculation of the shortcut fields. This robustness is not restricted to the $\Omega_0$ values used in the figure, but spans the whole range up to $2 \ \text{ps}^{-1} $. If $\Omega_0$ is further increased, then the performance is degraded. The degradation with increasing $\Omega_0$ is already obvious in the results obtained with model (\ref{bloch}) in Fig. \ref{fig:bloch}, since it overestimates the effect of phonons as pointed out above.
The purpose of using shortcut controls obtained with larger values of $\Omega_0$ is that they lead to appreciable shortcut detuning $\Delta(t)$, see Fig. \ref{fig:controls}, and the presence of this chirp in the shortcut pulses may be proven beneficial in the case where there are experimental imperfections, for example some uncertainty in the exciton transition frequency $\omega_0$. Thus, it is good to be assured that these robust chirped pulses perform well under the phonon influence on the system.

At this point it is worth to explain in more detail the physical origins of the difference between the results obtained with TEMPO and
Lindblad Eqs. (\ref{bloch}) in Fig. \ref{fig:TL}.
For all the cases depicted there we observe that for short durations there is a very good agreement between the two results. The reason is that, as we point out above, for such durations $\Omega(t)$ is actually a delta pulse which quickly accomplishes the desired transfer, so there is no much time left for the dissipative mechanisms to degrade the performance. As the pulse duration increases and dissipation finds more room to act, we observe a deviation between the results obtained with the two methodologies, but with a further increase in duration the good agreement is restored, at least for the lower temperature cases depicted in Figs. \ref{fig:LT1} and \ref{fig:LT10}. This behavior can be physically understood by considering the approximations under which Lindblad equations were derived, discussed above Eqs. (\ref{bloch}), which require that the control fields change slowly with time. Thus, for pulse durations which are long enough for the dissipative mechanisms to affect the system but at the same time short enough so the control variations are not slow relative to the timescale needed for the validity of Lindblad equations, we observe a difference between the two results. For longer pulse durations the difference is reduced, as in Fig. \ref{fig:LT1}, but as the temperature increases and the dissipative effects become more pronounced the difference persists for longer durations, as in Fig. \ref{fig:LT10}, while for large temperatures the substantial discrepancy indicates that the Lindblad model ceases to be valid, as shown in Fig. \ref{fig:LT50}.

\begin{figure}[t]
 \centering
\vspace{0cm}
 \begin{subfigure}[t!]{0.6\textwidth}
    \centering\caption{}\includegraphics[width=\linewidth]{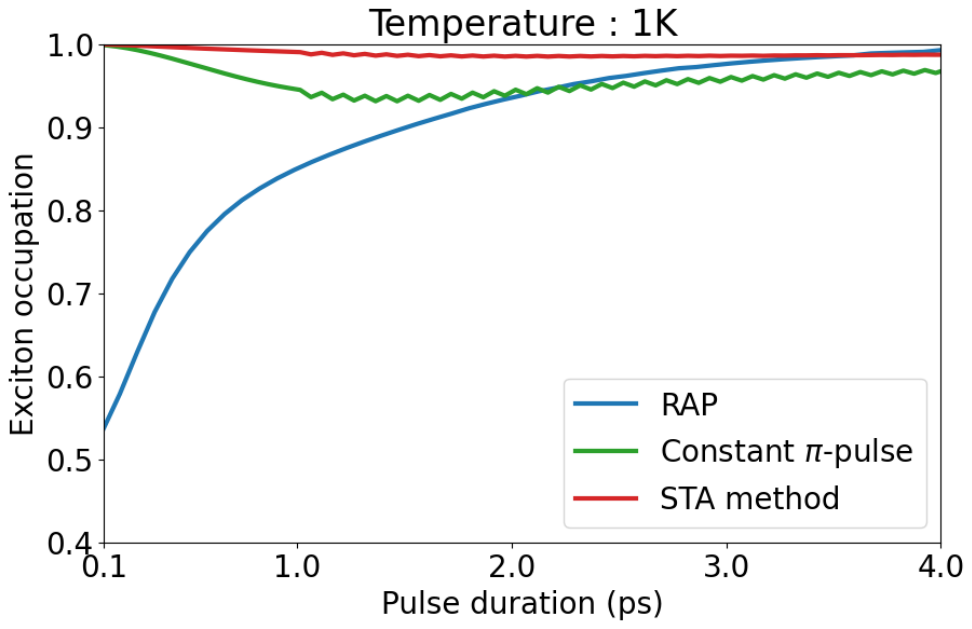}\label{fig:comp1}
\end{subfigure}  \\
\vspace{-7.5cm}
\begin{subfigure}[c]{0.6\textwidth}
    \centering\caption{}\includegraphics[width=\linewidth]{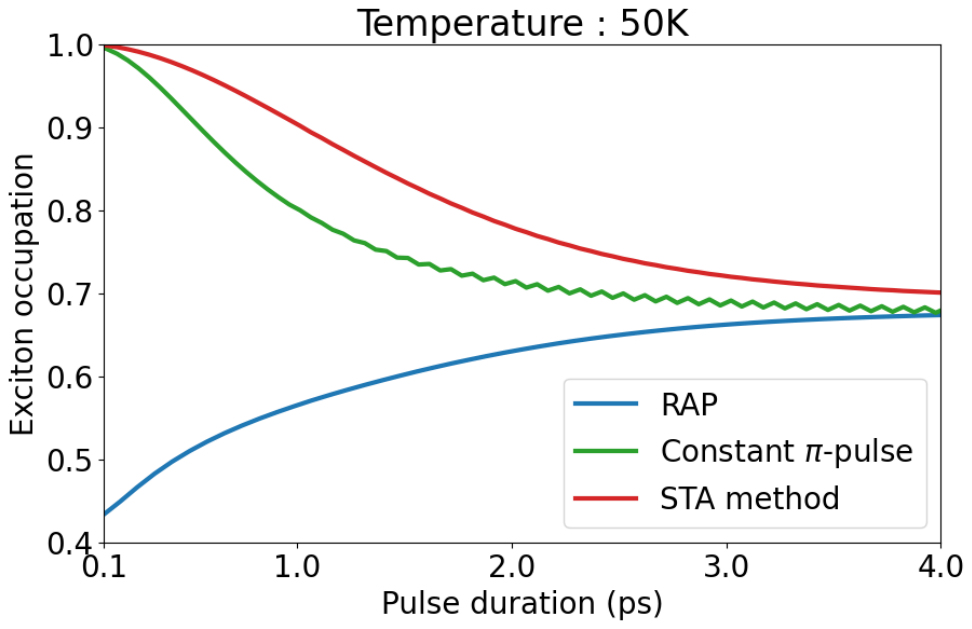}\label{fig:comp50}
\end{subfigure}
\vspace{-7cm}
\caption{Final exciton population calculated with TEMPO versus pulse duration, for the shortcut pulses (red curve), rapid adiabatic passage (blue curve), and a constant $\pi$-pulse (green curve), for temperatures $ 1 \ \text{K} $ (a) and $ 50 \ \text{K} $ (b).}
\label{fig:Comp}
\end{figure}

We finally compare the performance of the shortcut (STA) method with that of two other methods, a constant $\pi$-pulse with $\Omega(t) = \pi/T $ and $\Delta(t)=0$, and rapid adiabatic passage (RAP) implemented by a Gaussian pulse with linear chirp
\begin{equation}
\label{G_chirped}
\Omega(t) = \frac{\mathcal{A}}{\sqrt{2\pi T t_p}} \exp \left(  - \frac{t^2}{2 t_p^2} \right) \exp \left(-i\omega_0 t-\frac{ict^2}{2}  \right),
\end{equation}
which is obtained after passing a constant frequency Gaussian pulse with area $\mathcal{A}$ and initial duration $T$ (standard deviation) through a chirp filter with chirp constant $a$ \cite{Luker12}.
The pulse duration is modified from $T$ to \cite{Luker12}
\begin{equation}\label{eq:tpa}
t_p = \sqrt{T^2 + \frac{a^2}{T^2}}\,,
\end{equation}
while its frequency acquires a linear chirp
\begin{equation}
\label{chirp}
\Delta(t)=ct,
\end{equation}
with rate \cite{Luker12}
\begin{equation}\label{eq:ca}
c = \frac{a}{a^2+T^4}.
\end{equation}
For the chirped pulse we use the parameter values $\mathcal{A} = \pi$ and $a = 10 \ \text{ps}^{2}$, while the duration $T$ of the initial Gaussian pulse is used as a variable.
In Fig. \ref{fig:Comp} we show the final exciton population achieved by the three methods  (STA: red, RAP: blue, $\pi$-pulse: green) versus duration, for two different representative temperatures, $1 \ \text{K} $ and $50 \ \text{K} $. All the results are obtained using TEMPO. Observe that, although the $\pi$-pulse performs well for very short durations while RAP achieves good efficiency for longer durations, the shortcut method surpasses them in all the cases displayed, since it combines the advantages of both worlds (resonant pulses and adiabatic passage).

\section{Conclusion and future work}
\label{sec:conclusion}

In this work, we used pulses found with the STA method for the efficient creation of the exciton state in a GaAs/InGaAs quantum dot, under the acoustic phonon decoherence mechanism.
We propagated the system in time using the time-evolving matrix product operator (TEMPO) method and found that, for temperatures below $ 20 \ \text{K} $ and pulse duration up to $ 10 \ \text{ps} $, a very good transfer efficiency is achieved in general. We explained the results using a Bloch-like equation which sufficiently describes system dynamics at these lower temperatures. For higher temperatures, the transfer efficiency is substantially decreased except for subpicosecond pulses, where the shortcut Rabi frequency becomes basically a delta pulse accomplishing a fast inversion of population. This study is expected to find application in quantum technologies where quantum dots are used for single-photon generation on demand. A potential future extension of this work is to try to optimize the available controls (Rabi frequency and detuning) for the simplified Bloch equation, and then simulate the optimal controls found on the real system using the TEMPO method. This approach is computationally more efficient than optimizing directly with TEMPO and it may lead to near optimal controls, at least for the lower temperatures where the Bloch equation is a valid description of the system evolution. Of course, a complete investigation requires for comparison reasons to optimize also the exact dynamics obtained with TEMPO, which is a computationally demanding task, thus the study of optimal controls deserves a separate treatment.

\section*{Appendix - Details for TEMPO convergence}

The time evolution of the reduced density matrix of an open quantum system interacting with an environment is governed by the Feynman-Vernon Influence Functional (IF), which encapsulates the influence of the environment over the system, thereby introducing non-Markovianity. To accommodate for the finite memory, an Augmented Density Tensor (ADT) is constructed, which can be iteratively propagated in time using the discretized IF tensors. The TEMPO algorithm uses Singular Value Decomposition (SVD) to approximate a high-rank IF tensor into a chain of low-rank Matrix Product Operators (MPO), and similarly compress the ADT into a Matrix Product State (MPS) representation. This reduces the complexity, from exponential in memory length down to polynomial. The system evolves in three major steps. It begins with the growth phase, which lasts till the used memory length has been exhausted. The propagate phase follows and continues till the completion of the evolution. Here, the system is propagated forward in time using only the fixed amount of memory from the present instant and disregarding all the unnecessary past. Lastly, the fully grown tensor network is contracted row-wise and the final evolved state is extracted by tracing out all the indices of the final MPS except the one representing the current time.

The TEMPO algorithm uses three main computational parameters that determine the tradeoff between accuracy of the predicted dynamics and the time taken (or the computational complexity) for optimal convergence. The parameter \texttt{tcut} denotes the memory cut-off time, and must be long enough to capture all non-Markovian effects of the environment. The parameter \texttt{dt} is related to the time-step interval which must be small enough to provide sufficient resolution of the system dynamics and avoid Trotter error. And finally, the parameter \texttt{epsrel} determines the singular value cutoff threshold for numerical compression, which must be small enough to avoid significant data loss. While searching for the optimal parameter values to achieve convergence, one must take into account their interdependency and cannot continue the search process in isolation. We initiate the search space using OQuPy’s  \texttt{guess\_tempo\_parameters()} function \cite{OQuPy-paper,OQuPy-code-v0.5.0}, which provides only a rough estimate based on the time-dependent system frequencies and bath correlations. Using these as pivotal points we vary the parameter values within a suitable range and compute the system dynamics expectations ($\langle \hat{s}_x \rangle$, $\langle\hat{s} _y \rangle$, $\langle \hat{s}_z \rangle$) as a function of time for all of them. This allows us to further fine-tune upon the guessed parameters and finally reach the optimal values thereby establishing a proof of convergence for the given system-bath under investigation.

The problem in our case becomes more challenging because we are dealing with laser pulses of varied durations ranging from 0.01–10 ps, hence the TEMPO parameters need to be adjusted accordingly for different pulse durations to avoid unnecessary cost while also ensuring sufficient resolution. To deal with this issue, we start off by selecting \texttt{tcut}. Having called the \texttt{guess\_tempo\_parameters()} method using a tolerance of 0.005 for multiple pulse durations and also plotted the bath correlation dynamics, we concluded to the final value \texttt{tcut} = 2 ps. This was further confirmed by the plots of expectations obtained with larger values of \texttt{tcut} as well (which should obviously work but at greater costs) that were observed to superimpose on top of one another. This also means that for pulses with durations less than 2 ps, the finite memory approximation is not valid but since all those computations complete within a couple of seconds or less, therefore we don’t face any great disadvantage.

A similar procedure is carried out for both \texttt{dt} and \texttt{epsrel}, and we finally conclude that for pulses with duration in the range 0.01–0.1 ps, we shall use ($\texttt{dt}$, $\texttt{epsrel}$) = (0.001, $2 \times 10^{-7}$), for the range 0.1–2 ps ($\texttt{dt}$, $\texttt{epsrel}$) = (0.01, $2 \times 10^{-6}$), and for the rest 2–10 ps we choose ($\texttt{dt}$, $\texttt{epsrel}$) = (0.1, $2 \times 10^{-5}$). Now, since we are fixing \texttt{tcut}, we need to decrease \texttt{epsrel} while decreasing \texttt{dt} because at smaller time scales we need a finer resolution of system dynamics. One might have a natural urge to set the smallest possible \texttt{epsrel} value for all \texttt{dt}. Note also that as the time complexity of the singular value decompositions in the TEMPO tensor network scales with the third power of the internal bond dimension, which is directly controlled by the precision, the computational time increases very rapidly.

\begin
  {figure}[t]
    \centering
     \begin{subfigure}[t!]{0.45\textwidth}
    \centering\caption{}\includegraphics[width=\linewidth]{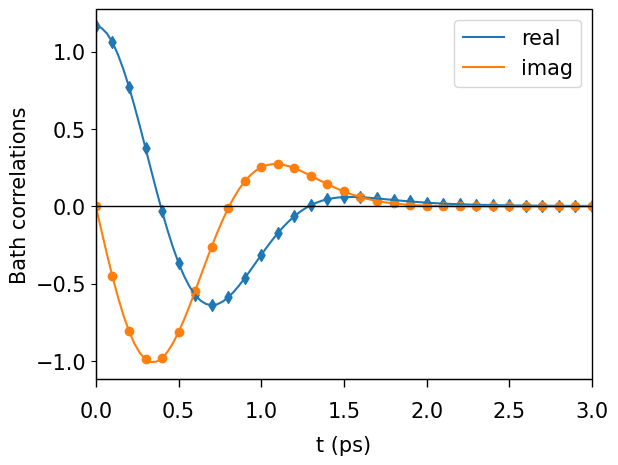}\label{fig:corr_decay}
    \end{subfigure} \\
     \begin{subfigure}[t!]{1.0\textwidth}
    \centering\caption{}\includegraphics[width=\linewidth]{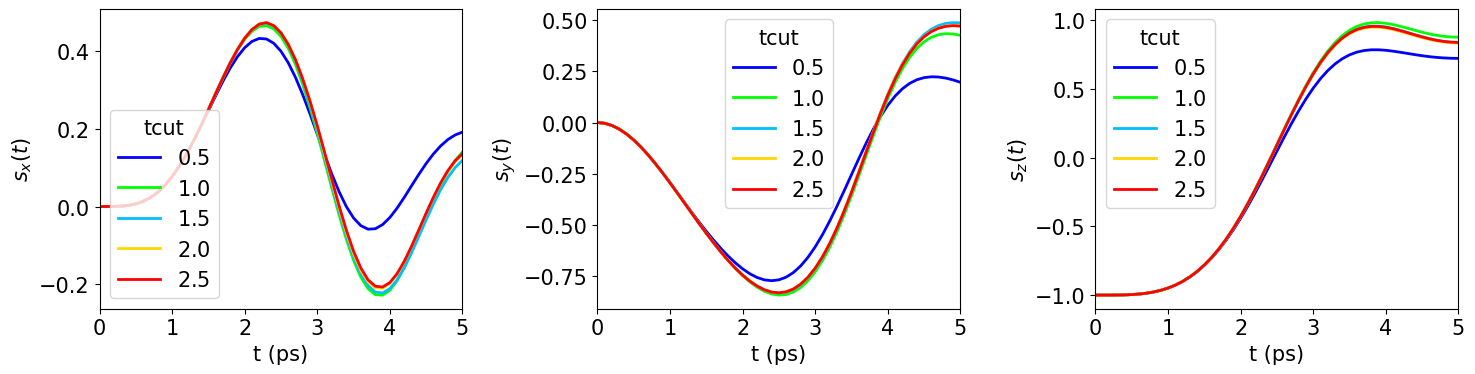}\label{fig:tcut}
\end{subfigure} \\
     \begin{subfigure}[t!]{1.0\textwidth}
    \centering\caption{}\includegraphics[width=\linewidth]{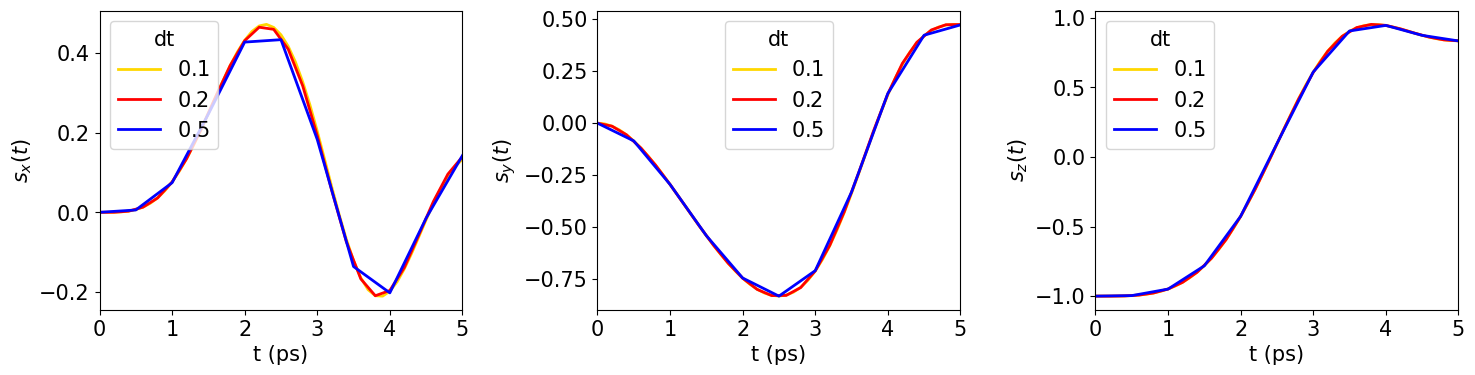}\label{fig:dt}
\end{subfigure}
     \begin{subfigure}[t!]{1.0\textwidth}
    \centering\caption{}\includegraphics[width=\linewidth]{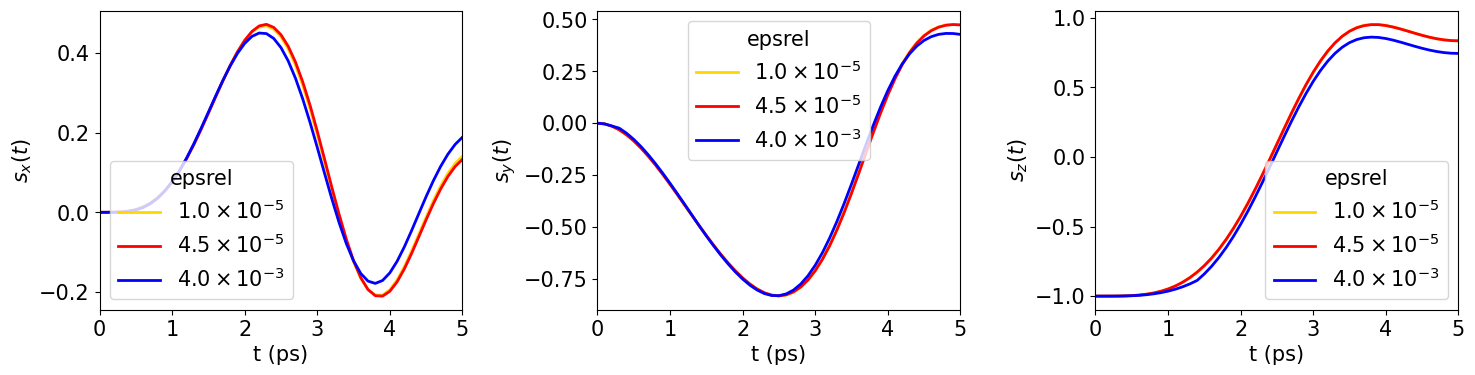}\label{fig:epsrel}
\end{subfigure}
    \caption{(a) Decay of the bath correlation function with time. Convergence of the time evolution of the Bloch vector components $(s_x, s_y, s_z)$ when varying the TEMPO parameters (b) \texttt{tcut} in the range \{0.5, 1.0, 1.5, 2.0, 2.5\} ps, (c) \texttt{dt} in the range \{0.1, 0.2, 0.5\} ps, and (d) \texttt{epsrel} in the range \{$1.0\times10^{-5}$, $4.5\times10^{-5}$, $4.0\times10^{-3}$\}. All the plots are obtained for STA pulses of duration $T = 5$ ps, Rabi frequency amplitude $6\pi/T \ \text{ps}^{-1}$ and at bath temperature of 1K.}
    \label{fig:tempo_params}
\end{figure}

\begin{figure}[t]
    \centering
    \includegraphics[width=0.5\linewidth]{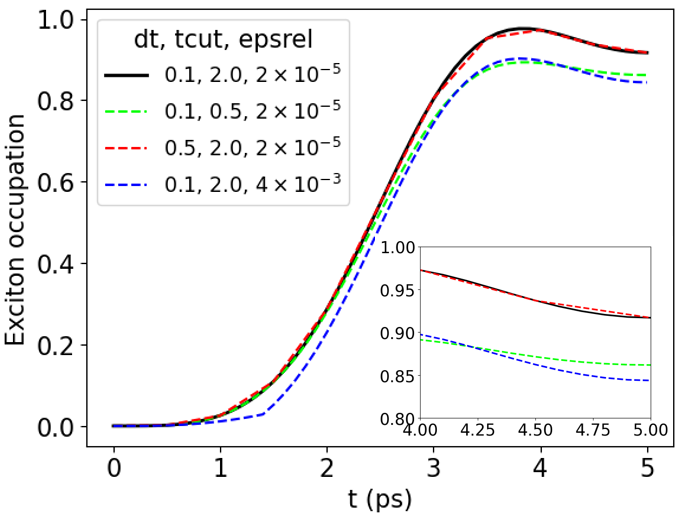}
    \caption{Population dynamics for a quantum dot driven by STA pulses of duration $T = 5$ ps, Rabi frequency amplitude $6 \pi/T$ $\text{ps}^{-1}$ and at bath temperature of 1K. Comparison between the result obtained using optimal TEMPO parameters (black solid line) and those obtained with three sets of different combinations represented by the dashed lines. In each set, a single TEMPO parameter has been distorted from the optimal value, keeping the other two fixed. Distortions in tcut, dt and epsrel from the optimal values result in the green, red and blue dashed lines, respectively.}
    \label{fig:conv_comp}
\end{figure}

We shall demonstrate the search for these optimal parameters using STA pulses of duration (T) 5 ps and Rabi frequency amplitude $6\pi/T \ \text{ps}^{-1}$ at a bath temperature of 1K. The initial guess suggests \texttt{dt} = 0.125 ps, \texttt{tcut} = 2.125 ps, and \texttt{epsrel} = $4.521\times10^{-5}$. The easiest way to verify \texttt{tcut} is by plotting the correlation function versus time using the built-in function \texttt{oqupy.helpers.plot\_correlations\_with\_parameters()}. Fig. \ref{fig:corr_decay} shows that the bath correlations have almost decayed at 2 ps, and therefore we can set \texttt{tcut} = 2 ps. For a more detailed analysis, we plot the expectation values for different choices of \texttt{tcut} = \{0.5, 1.0, 1.5, 2.0, 2.5\} ps in Fig. \ref{fig:tcut}. We observe that the curves corresponding to 0.5 ps (blue) and 1.0 ps (green) diverge from the others. While the choice 1.5 ps (cyan curve) seems to be a good fit when considering the correlations plot  \ref{fig:corr_decay}, it also diverges at some points. The yellow curve corresponding to 2 ps converges with the red curve corresponding to 2.5 ps, indicating that 2 ps is a successful choice that allows the bath correlations to decay significantly. Next, we plot in Fig. \ref{fig:dt} the expectations using different values of \texttt{dt} = \{0.1, 0.2, 0.5\} ps with the same \texttt{epsrel} = $4.521\times10^{-5}$. 
We observe that as \texttt{dt} is reduced the plots converge and we choose \texttt{ dt} = 0.1.
Finally, using the obtained parameters we vary \texttt{epsrel} = \{$1.0\times10^{-5}$, $4.5\times10^{-5}$, $4.0\times10^{-3}$\} and plot the expectation values in Fig. \ref{fig:epsrel}. Observe that, while the blue curves corresponding to the larger value diverge, the red and yellow curves corresponding to the smaller values are indistinguishable, confirming the convergence of the dynamics. As stated above, we shall use \texttt{epsrel} = $2\times10^{-5}$ which applies to all pulse durations in the range of 2-10 ps. Fig. \ref{fig:conv_comp} compares the dynamics generated by the optimal parameters (solid black curve) and the non-optimal ones (dashed curves), where we distort one of the parameters from its optimal value in each of the three cases shown.

Finally, we have also tested the successful operability of these parameters for different bath temperatures in the range 0–100~K with \texttt{omega\_cutoff} of 2~meV and $A = 0.0112$~ps/K by comparing with results from the literature, for example, results from Refs. \cite{Cutcheon10,Luker12}.

\end{document}